\documentclass[aps,prl,showpacs,amsmath,amssymb,onecolumn]{revtex4-1}
\usepackage{epsfig}
\usepackage{graphicx,color}
\usepackage{amsmath}
\newcommand{\beq}{\begin{eqnarray}}
\newcommand{\eeq}{\end{eqnarray}}

\begin{document}

\title{Theory of inclusive breakup cross section for Borromean nuclei within a four-body spectator model}
\author{B. V. Carlson}
\affiliation{ Instituto Tecnol\'{o}gico de Aeron\'{a}utica, DCTA,12.228-900 S\~{a}o Jos\'{e} dos Campos, SP, Brazil }
\author{ T. Frederico}
\affiliation{ Instituto Tecnol\'{o}gico de Aeron\'{a}utica, DCTA,12.228-900 S\~{a}o Jos\'{e} dos Campos, SP, Brazil}
\author{M. S. Hussein}
\affiliation{Instituto Tecnol\'{o}gico de Aeron\'{a}utica, DCTA,12.228-900 S\~{a}o Jos\'{e} dos Campos, SP, Brazil}
\affiliation{Instituto de Estudos Avan\c{c}ados, Universidade de S\~{a}o Paulo C. P.
72012, 05508-970 S\~{a}o Paulo-SP, Brazil}
\affiliation{Instituto de F\'{\i}sica,
Universidade de S\~{a}o Paulo, C. P. 66318, 05314-970 S\~{a}o Paulo,-SP,
Brazil}

\keywords{Inclusive breakup, heavy-ion scattering, Borromean nuclei}
\pacs{24.10Eq, 25.70.Bc, 25.60Gc }

\begin{abstract}
We develop a model to treat the inclusive non-elastic break up reactions involving weakly bound 
three-cluster nuclei. Borromean, two-nucleon, halo nuclei are candidates of unstable three-fragments 
projectiles. The model is based on the theory of inclusive breakup reactions commonly employed in 
the treatment of incomplete fusion and surrogate method. The theory was developed in the 80's by 
Ichimura, Autern and Vincent (IAV) [Phys. Rev. C 32, 431 (1985)] \cite{IAV1985},  
Udagawa and Tamura (UT)[Phys. Rev. C 24, 1348 (1981)], \cite{UT1981} and Hussein and McVoy (HM)[Nucl. Phys. A 445, 
124 (1985)], \cite{HM1985}. We extend these three-body theories to derive an expression for the fragment 
yield in the reaction $A\,(a,b)\,X$, where the projectile is $a = x_1 + x_2 + b$. 
The inclusive breakup cross section is found to be the sum of a generalized four-body form of the 
elastic breakup cross section plus the inclusive non-elastic breakup cross section which involves 
the "reaction" cross section of the participant fragments, $x_1$ and $x_2$. The final result is 
similar to the three-body case reviewed in Austern, et al. (Phys. Rep. \textbf{154}, 125 (1987)), \cite{Austern1987}, 
but with important genuine four-body effects added, both in the elastic breakup cross section, which 
now contains the full correlations between the participant fragments,  and in the inclusive non-
elastic breakup. These developments should encourage experimentalists to seek more information about 
the $x_{1} + x_{2}$ system in the elastic breakup cross section, and to theorists to further develop and 
extend the surrogate method, based on the inclusive non-elastic breakup part of the $b$ spectrum.
\end{abstract}

\maketitle

\section{ Introduction} 
Recently interest in extracting the neutron capture cross section by stable nuclei at higher 
energies through the $(d, p)$ reaction has arisen in part for application to next generation 
reactors (Fast Breeder Reactors fueled by $^{238}$U, and $^{232}$Th), and in part for the study of 
the reaction mechanism of weakly bound stable nuclei. There is also potential 
application to the production of elements in the r-process of nucleosynthesis \cite{Ducasse2015}. In a recent 
publication \cite{Potel2015}, tested the Surrogate Method \cite{Escher2012} in the case of $(d, p)$ 
reaction on the actinide nuclei to be used in these projected reactors. For this purpose, they 
employed the theory of inclusive non-elastic breakup reactions, where the proton is treated as a 
spectator, merely scattering off the target, and the neutron is captured by the target, and, at 
higher energies, inelastically scattering from the target. Other papers on the $(d, p)$
reaction were also published in 2015 dealing with the same issue \cite{Moro2-2015,Carlson2015}. 
Ref. \cite{Moro2-2015} also discussed the application of this  hybrid picture (direct breakup 
followed by compound nucleus formation of the subsystem)  to the reaction $^{6}$Li +$^{209}$Bi 
$\rightarrow \alpha + X$, at $E_{Lab.}$ = 24 MeV and 32 MeV. The theory employed in all these 
publications was developed in the 80's \cite{IAV1985, HM1985, UT1981}. The inclusive non-elastic 
breakup part of the cross section has come to be known as the Austern formula which involves the 
reaction cross section of the "captured" fragment, calculated with a full three-body scattering wave 
function \cite {Austern1987}. At much higher energy deuteron or other breaking projectiles, 
researchers relied on the very simple but physically transparent Serber model \cite{Serber1950}, 
which is a natural limiting approximation of the Austern formula. So far, no attempt has been made 
to apply the hybrid theory to the extraction of the neutron capture cross section involving 
radioactive nuclei. 

At near-barrier energies the three-body Austern formula alluded to above can in principle calculate 
the incomplete fusion part of the total fusion cross section. Extension of the three-body Austern 
form of the inclusive breakup cross section to the case of reactions induced by secondary beams of 
three-fragments projectiles, such as the weakly bound stable nucleus $^9$Be = $\alpha + \alpha$ + n, 
and Borromean, two-neutron halo, nuclei like $^6$He = $\alpha$ + n + n, $^{11}$Li = $^9$Li + n + n, 
$^{14}$Be = $^{12}$Be + n + n, and $^{22}$C = $^{20}$C + n + n, is certainly important as more data 
on these reactions have become available. Data on complete fusion and total fusion around the 
Coulomb barrier are currently being obtained and analyzed using the effective two-body "four-body" 
Continuum Discretized Coupled Channels model, which is basically unable to calculate the incomplete 
fusion part of the total fusion cross section \cite{DTT2002}, requiring urgent derivation and 
developments of the the four-body inclusive breakup cross section. This is the purpose of this 
paper. \\

\section{Derivation of the 4-body Austern formula} 
In the following, we supply the full details of the derivation of the exact four-body formula for inclusive non-elastic breakup cross section,
We write the  full Hamiltonian of the system, $x_1 + x_2 + b + A$ as,
\begin{equation}
H = K_{x_1} + K_{x_2} + K_b + V_{b{x_1}} + V_{b{x_2}}+ V_{{x_1}A} + V_{{x_2}A}+ V_{bA} ++ h_b + h_{x_1} + h_{x_2} +  h_{A} 
\end{equation}
where the $h$'s denote the intrinsic Hamiltonians, $K_i$ the kinetic energy operator of fragment $i$, and The $V$'s are the microscopic interaction between pairs of fragments.

To proceed with the derivation we make the following important approximations:

\begin{enumerate}

\item
The fragments, $x_1$, $x_2$ and $b$ are structureless,\\
\begin{equation}
h _b = h_{x_1} = h_{x_2} = 0
\end{equation}

The interaction $V_{bA}$ is replaced by the complex optical potential $U_{bA}$. This corresponds to treating $b$ as a spectator.\\

\item
The target is much more massive compared to the projectile and one sets $K_A = 0$ ( the adiabatic assumption). \\

\end{enumerate}

We call the reduced Hamiltonian of the system, $H_{sm}$, where sm stands for "spectator model".
\begin{equation}
H_{sm} = K_{x_1} + K_{x_2} + K_b + V_{b{x_1}} + V_{b{x_2}} + V_{{x_1} {x_2}} + V_{{x_1}A} + V_{{x_2}A} +  U_{b} + h_{A}
\end{equation}

The still many-body Schr\"{o}dinger equation is now,
\begin{equation}
H_{sm}\varXi(\textbf{r}_{x_1}, \textbf{r}_{x_2}, \textbf{r}_b, A) = E \varXi(\textbf{r}_{x_1}, \textbf{r}_{x_2}, \textbf{r}_b, A)
\end{equation}
where the argument $A$ inside the many-body wave function $\varXi(\textbf{r}_{x_1}, \textbf{r}_{x_2}, \textbf{r}_b, A)$ stands for the A internal coordinates of the target nucleus.
The exact wave function of the $x_1 + x_2 + A$ system is denoted by $\varPsi^{c}_{xA}$, with $c$ corresponding to a state in this system. This state can be bound or unbound (continuum or scattering). The three-body eigenvalue equation which describes the $x_1 + x_2 + A$ system is,
\begin{equation}\label{xA}
[K_{x_1} + K_{x_2} + V_{{x_1}A} + V_{{x_2}A} + V_{x_{1}x_{2}}+ h_A]\varPsi^{c}_{xA} \equiv H_{xA}\varPsi^{c}_{xA} = E^c\varPsi^{c}_{xA}
\end{equation}

We are now in a position to calculate the inclusive breakup cross section. The process is $ a + A 
\rightarrow b + (x_1 + x_2 + A)$, and the cross section to be calculated is the double differential 
cross section for observing the spectator fragment, $b$, $d^2\sigma_{b}/(dE_{b}d\Omega_b)$
\begin{equation}
\frac{d^2\sigma_b}{dE_{b}d\Omega_b} = \frac{2\pi}{\hbar v_a}\rho_{b}(E_b)\sum_{c} \left|
\left\langle\chi_{b}^{(-)}(\textbf{r}_{b})\varPsi^{c}_{xA}\left|V_{bX}\right|\varXi(\textbf{r}_{x_1},
\textbf{r}_{x_2}, \textbf{r}_b, A) \right\rangle \right|^{2} \delta(E - E_b - E^c) 
\label{cross-exact}
\end{equation}
where $V_{bX} \equiv V_{bx_1}+V_{bx_2}$ and $\chi_{b}^{(-)}(\text{r}_b)$ is the distorted wave of the spectator fragment, 
\begin{equation}
\left[E_b - U^{\dagger}_{b} - K_b\right]\left|\chi_{b}^{(-)}(\text{r}_b)\right\rangle = 0
\end{equation}
In the above equation, the complex conjugate of the optical potential operator appears, owing to the 
fact that the distorted wave being calculated is the one with outgoing wave boundary condition. If 
the solution required is $\chi_{b}^{(+)}(\text{r}_b)$, then the equation becomes $[E_b - U_{b} - 
K_{b}]\chi_{b}^{(+)}(\text{r}_b) = 0$. The density of states of the 
$b$ fragment is given by, $\rho_{b}(E_b) = \mu_{b}k_{b}/ [(2\pi)^{3}\hbar^2]$, is just the result of 
the change of variables: $d\textbf{k}_{b} /(2\pi)^3= k_{b}^2 dk_{b} d\Omega_{b}/(2\pi)^3 = \mu_b 
k_{b}/[(2\pi)^3\hbar^2] dE_{b} d\Omega_{b} = \rho_{b}(E_b) dE_{b}d\Omega_{b}$. The triple 
differential cross section is given by $d^3\sigma_{b}/[d\textbf{k}_{b}/(2\pi)^3]$ and it is just the 
formulae above without the density of states factor.

The exact spectator inclusive breakup cross section, Eq. (\ref{cross-exact}), can be reduced to the following form,
\begin{align}
\frac{d^{2}\sigma_b}{dE_{b}d\Omega_{b}} 
= -\frac{2}{\hbar v_a} \rho_{b}(E_b) Im\left[\left\langle \varXi\left|V_{bX}\right|
\chi_{b}^{(-)}\right\rangle G^{(+)}_{{x_1}{x_2}A}\left\langle \chi_{b}^{(-)}\left|V_{bX}\right|
\varXi \right\rangle\right]
\label{cross-Green}
\end{align}
where $G^{(+)}_{{x_1}{x_2}A}$ is the $x_1 + x_2 + A$ Green's function, given by,
\begin{equation}
G^{(+)}_{{x_1}{x_2}A} = \frac{1}{E - E_b - K_{x_1} - K_{x_2} - V_{{x_1}A} - V_{{x_2}A} - V_{{x_1}{x_2}}+ i\varepsilon} = \frac{1}{E -E_b - H_{{x_1}{x_2}A} + i\varepsilon} \label{Goperator}
\end{equation}
In all he above equations, $E^c=E-E_b$ is shorthand notation for the total energy of the $x_1 + x_2 + A$ system, as defined in Eq.(\ref{xA}). 
In deriving the above equation we have used several identities: $\delta(E -E_b -E^{c}) = -(1/\pi)Im[E- E_b -E^{c} + i\varepsilon]^{-1}$. Then, $[E - E_b -E^c + i\varepsilon]^{-1} \langle\varPsi^{c}_{xA}| = \langle \varPsi^{c}_{xA}| [E - E_b - H_{{x_1}{x_2}A}]^{-1} = \langle \varPsi^{c}_{{x_1}{x_2}A}|G^{(+)}_{{x_2}{x_2}A}(E^c)$. Once we remove the $c$ dependence from the Green's function we can  use closure to perform the sum over $c$. Eq. (\ref{cross-Green}), is not very useful since the wave function $|\varXi\rangle$ is a many-body wave function and the process at hand is basically a three-body one. Therefore, we use the approximation $|\varXi \rangle \approx |\Psi_{0}^{4B(+)}\Phi_{A}^{0}\rangle$, where 
$|\Psi_{0}^{4B(+)}\rangle$ is the exact four-body scattering wave function, and $|\Phi^{0}_{A}\rangle$ is the ground state wave function of the target. The cross section then acquires the form,
\begin{equation}\label{cross-Green-1}
\frac{d^{2}\sigma_b}{dE_{b}d\Omega_{b}} =  -\frac{2}{\hbar v_a} \rho_{b}(E_b) Im\left[\langle \Psi_{0}^{4B(+)}\left|V_{bX} \right|\chi_{b}^{(-)}\right\rangle \langle \Phi^{0}_{A}|G^{(+)}_{{x_1}{x_2}A}|\Phi^{0}_{A}\rangle \left\langle \chi_{b}^{(-)}\left|V_{bX} \right|\Psi_{0}^{4B(+)}\rangle\right] 
\end{equation}

The target ground state expectation value of the three-body Green's function, $\langle \Psi^{0}_{A}|G^{(+)}_{x_1x_2A}|\Phi^{0}_{A}\rangle$ requires a special attention. Since $V_{{x_1}{x_2}}$ does not have any reference to the target degrees of freedom, we can just lump it to the energy $E_x$, and perform the average using the usual projection operator techniques. The result is
\begin{equation}
\langle \Phi^{0}_{A}|G^{(+)}_{{x_1}{x_2}A}|\Phi^{0}_{A}\rangle = \frac{1}{E_x - K_{{x_1}} - K_{{x_2}} - V_{{x_1}{x_2}} - U_{{x_1}{x_2}}+i\varepsilon}  \equiv G^{(+)}_{{x_1}{x_2}}
\label{Goptical}
\end{equation} 
where we have defined the complex optical potential of $x_1 + x_2$ system, $U_{{x_1}{x_2}}$ as 
\begin{equation}
U_{{x_1}{x_2}} = U_{x_1} + U_{x_2} + U_{3B} \label{OP}
\end{equation}
The $U_{x}$'s are the complex optical potentials of the $x_1$ and $x_2$ fragments and $U_{3B}$ is a three-body complex (optical) average potential that describes processes which involve the excitation of the target by one fragment and its de-excitation by the other fragment.
 
We now derive the imaginary part of the A-averaged Green's function of Eq. (\ref{Goptical}), $G^{(+)}_{{x_1}{x_2}}$.  First we re-write this Green's function as
\begin{equation}
G^{(+)}_{{x_1}{x_2}} = \frac{1}{E_x - H_{{x_1}{x_2}} - U_{{x_1}{x_2}} + i\varepsilon} = \frac{1}{E_x - H_0 - (V_{{x_1}{x_2}} + U_{{x_1}{x_2}}) + i\varepsilon}
\end{equation}
where $H_{{x_1}{x_2}} \equiv K_{x_1} + K_{x_2} + V_{{x_1}{x_2}} = H_0 + V_{{x_1}{x_2}}$ is Hermitian. Accordingly, the full potential, $V_{{x_1}{x_2}} + U_{{x_1}{x_2}}$ in the average, optical, Green's function is the sum of a real $x_1 + x_2$ interaction which generates correlations among these two fragments, plus the three-body optical potential of these fragments, Eq.(\ref{OP}), which we re-write $U = ReU - iW$. To simplify the notation, we call the interaction $V_X\equiv [Re\,U_X + V_X] - iW_{X}$, where now $X \equiv {x_1}{x_2}$. We now use an identity derived by several authors \cite{Ichimura1982, Hussein1984}
\begin{equation}
Im G^{(+)}_X\, =\,
-\pi\,\Omega^{(-)}_X\,\delta(E_x - H_0)\,
(\Omega^{(-)}_X)^\dagger - 
(G^{(+)}_X)^\dagger\, W_X\,G^{(+)}_X 
\label{ImaG}
\end{equation}
where the M\"oller operator $\Omega^{(-)}_{X} = [1 + G^{(-)}_{X}({V}_{X})^{\dagger}]$. When operating 
on a plane wave $|\mathbf{k}_X\rangle$ it generates a "three-body" distorted wave $\left|
\psi^{3B(-)}_{\mathbf{k}_{X}}\right\rangle$. The fragments, $x_1$ and $x_2$ are scattered by the target and are 
interacting with each other through the two-body potential $V_{X}$. They are also distorted by the 
"three-body" optical potential which also generates correlation. Thus, the total correlating 
interaction between the two unobserved fragments should be taken as $V_{X} + U_{3B}$. 

The final result for the 4-body inclusive breakup cross section is, from Eq.(\ref{cross-Green-1}),
\begin{multline}
\frac{d^{2}\sigma_b}{dE_{b}d\Omega_{b}}  = 
\frac{2}{\hbar v_a} \rho_{b}(E_b)\left[\langle \Psi_{0}^{4B(+)}|
V_{bX} |\chi_{b}^{(-)}\rangle
\sum_{{\textbf{k}_{X}}}\Omega^{(-)}_{X}
|\textbf{k}_{X} \rangle\, \delta(E_{x} - E_{{\textbf{k}_{X}}})\, \langle\textbf{k}_{X}|
(\Omega^{(-)}_{X})^{\dagger} \langle \chi_{b}^{(-)}\left|V_{bX} \right|\Psi_{0}^{4B(+)}\rangle + \right.\\ 
\left. + \frac{1}{\pi}\langle\Psi_{0}^{4B(+)}|
V_{bX} |\chi_{b}^{(-)}\rangle
(G^{(+)}_{X})^{\dagger}\,W_{X}\,G^{(+)}_{X}\langle\chi_{b}^{(-)}|
V_{bX} |\Psi_{0}^{4B(+)}\rangle\right]
\end{multline}
The cross section then becomes,
\begin{multline}
\frac{d^{2}\sigma_b}{dE_{b}d\Omega_{b}}  =  \frac{2\pi}{\hbar v_{a}}\rho_{b}
(E_b)\,\sum_{{\textbf{k}_{X}}} \left|\langle \chi^{3B(-)}_{X}\chi^{(-)}_{b}|V_{bX} |\Psi_{0}^{4B(+)}\rangle \right|^2 \delta(E - E_{b} - E_{{\textbf{k}_{X}}}) +
\\ 
 + \frac{2}{\hbar v_a}\rho_{b}(E_b)\left[\langle\Psi_{0}^{4B(+)}|V_{bX}
(G^{(+)}_{X})^{\dagger}|\chi_{b}^{(-)}\rangle[W_{x_1} + W_{x_2} + W_{3B}]\langle\chi_{b}^{(-)}|
G^{(+)}_{X}V_{bX}|\Psi_{0}^{4B(+)}\rangle\right]
\end{multline}
 We now write 
 $V_{b{x_1}} + V_{b{x_2}} 
 = [V_{b{x_1}}+ V_{b{x_2}} + H_{0} + V_{X} + U_{X} -E] - 
 [H_{0} + V_{X} +U_{X} - E]$. 
 Then the matrix element $\langle\Psi_{0}^{4B(+)}|[V_{b{x_1}} + 
 V_{b{x_2}}](G^{(+)}_{X})^{\dagger}|\chi_{b}^{(-)}\rangle$ becomes -$\langle\Psi_{0}^{4B(+)}|[H_{0} 
 + V_{X} + U_{X} - E] (G^{(+)}_{X})^{\dagger}|\chi_{b}^{(-)}\rangle = 
\langle\Psi_{0}^{4B(+)}|\chi_{b}^{(-)})$. Accordingly, the inclusive breakup cross section becomes,
\begin{equation}
\frac{d^{2}\sigma_b}{dE_{b}d\Omega_{b}} = \frac{d^{2}\sigma^{EB}_b}{dE_{b}d\Omega_{b}} + \frac{d^{2}\sigma^{INEB}_b}{dE_{b}d\Omega_{b}} 
\end{equation}
where the four-body elastic breakup cross section is,
\begin{equation}
\frac{d^{2}\sigma^{EB}_b}{dE_{b}d\Omega_{b}} =  \frac{2\pi}{\hbar v_{a}}\rho_{b}(E_b)\,\sum_{{\textbf{k}_{X}}} \left|\langle \chi^{3B(-)}_{X}\chi^{(-)}_{b}|[V_{{bx_{1}}} + V_{{bx_{2}}}]|\Psi_{0}^{4B(+)}\rangle \right|^2 \delta(E - E_{b} - E_{{\textbf{k}_{X}}}) \label{4BEB}
\end{equation}
and
\begin{equation}
\frac{d^{2}\sigma^{INEB}_b}{dE_{b}d\Omega_{b}} = \frac{2}{\hbar v_a}\rho_{b}(E_b) \langle \hat{\rho}_{X}|(W_{x_1} + W_{x_2} + W_{3B})|\hat{\rho}_{X}\rangle \label{CFH}
\end{equation}
with the source function 

\begin{equation}
\hat{\rho}_{X}(\textbf{r}_{x_{1}},\textbf{r}_{x_{2}}) = (\chi_{b}^{(-)}|\Psi_{0}^{4B(+)}\rangle = \int d\textbf{r}_{b}(\chi_{b}^{(-)}(\textbf{r}_{b}))^{\dagger}\Psi_{0}^{4B(+)}(\textbf{r}_{b}, \textbf{r}_{x_{1}}, \textbf{r}_{x_{2}})\label{rho-4B}
\end{equation}

depending only on the coordinates of $x_1$ and $x_2$.\\

\section{Analysis of the 4B inclusive breakup cross section}
The above equations, Eqs.(\ref{4BEB}, \ref{CFH}) are the main result of this paper. They supply the spectrum of a fragment in the breakup of a three-fragment projectile. They represent an important non-trivial generalization of the well known case of the breakup of a two-fragment projectile. The elastic breakup formula involves the three fragments in the continuum, two of which are not observed, and they are correlated. It would be very interesting to perform  a measurement of coincidence of $b$ and, say, $x_1$ or better yet a three fragment coincidence of the three fragments to learn about the correlations between the participant $x_1$ and $x_2$. \\

The second equation, Eq.(\ref{CFH}), is the four-body inclusive non-elastic breakup cross section. It 
differs significantly from the three-body Austern formula \cite{Austern1987}. We refer to Eq.(\ref{CFH}) as the Carlson-
Frederico-Hussein (CFH) formula. The major new features present can be quantified by writing the 
formula as a sum of three terms,
\begin{equation}
\frac{d^{2}\sigma^{INEB}_b}{dE_{b}d\Omega_{b}} = \frac{k_a}{E_a}\left[\frac{{E_{x_1}}}
{{k_{x_1}}}\sigma_{R}^{x_1} +\frac{{E_{x_2}}}{{k_{x_2}}}\sigma_{R}^{x_2} + \sqrt{\frac{{E_{x_1}}}
{{k_{x_1}}} \frac{{E_{x_2}}}{{k_{x_2}}}}\sigma_{R}^{3B}\right] \label{CFH-R}
\end{equation}
where, 
\begin{equation}
\sigma_{R}^{x_1} = \frac{{k_{x_1}}}{E_{{x_1}}} \langle \hat{\rho}_{{x}_1, {x}_2}|W_{x_1}|\hat{\rho}_{{x}_1, {x}_2}\rangle, \label{sigma_1}
\end{equation}
\begin{equation}
\sigma_{R}^{x_2} = \frac{{k_{x_2}}}{E_{{x_2}}} \langle \hat{\rho}_{{x}_1, {x}_2}|W_{x_2}|\hat{\rho}_{{x}_1, {x}_2}\rangle, \label{sigma_2}
\end{equation}
and,
\begin{equation}
\sigma_{R}^{3B} =  \sqrt{\frac{{k_{x_1}}}{{E_{x_1}}} \frac{{k_{x_2}}}{{E_{x_2}}}}\langle \hat{\rho}_{{x}_1, {x}_2}|W_{3B}|\hat{\rho}_{{x}_1, {x}_2}\rangle \label{sigma_3B}
\end{equation}

The cross section, $\sigma_{R}^{x_1}$ represents the absorption of fragment $x_1$ by the target, while fragment
$x_2$ just scatters off the target through the optical potential $U_{{x}_1 A}$. The second cross section,  $\sigma_{R}^{x_2}$ is just the exchange of the role of these 
two fragments; fragment $x_2$ is captured by the target and fragment $x_1$ is scattered. Note that these cross sections are different from the one which appears in the three-body theory. The three -body sub-system $x_{1} -x_{2} -A$ is not treated within the spectator model, while the b- x -A system is. Thus we anticipate that, say,  $\sigma_{R}^{{x}_1}$ in a (t,p)
reaction, to be different from $\sigma_{R}^{x_1}$ extracted from a (d,p) reaction. We return to this point later.
Finally the last cross section, $\sigma_{R}^{3B}$, is new and a genuine three-body absorption cross section. In the following we take a critical look at its structure.

\begin{figure}[htb]
\centerline{\epsfig{figure=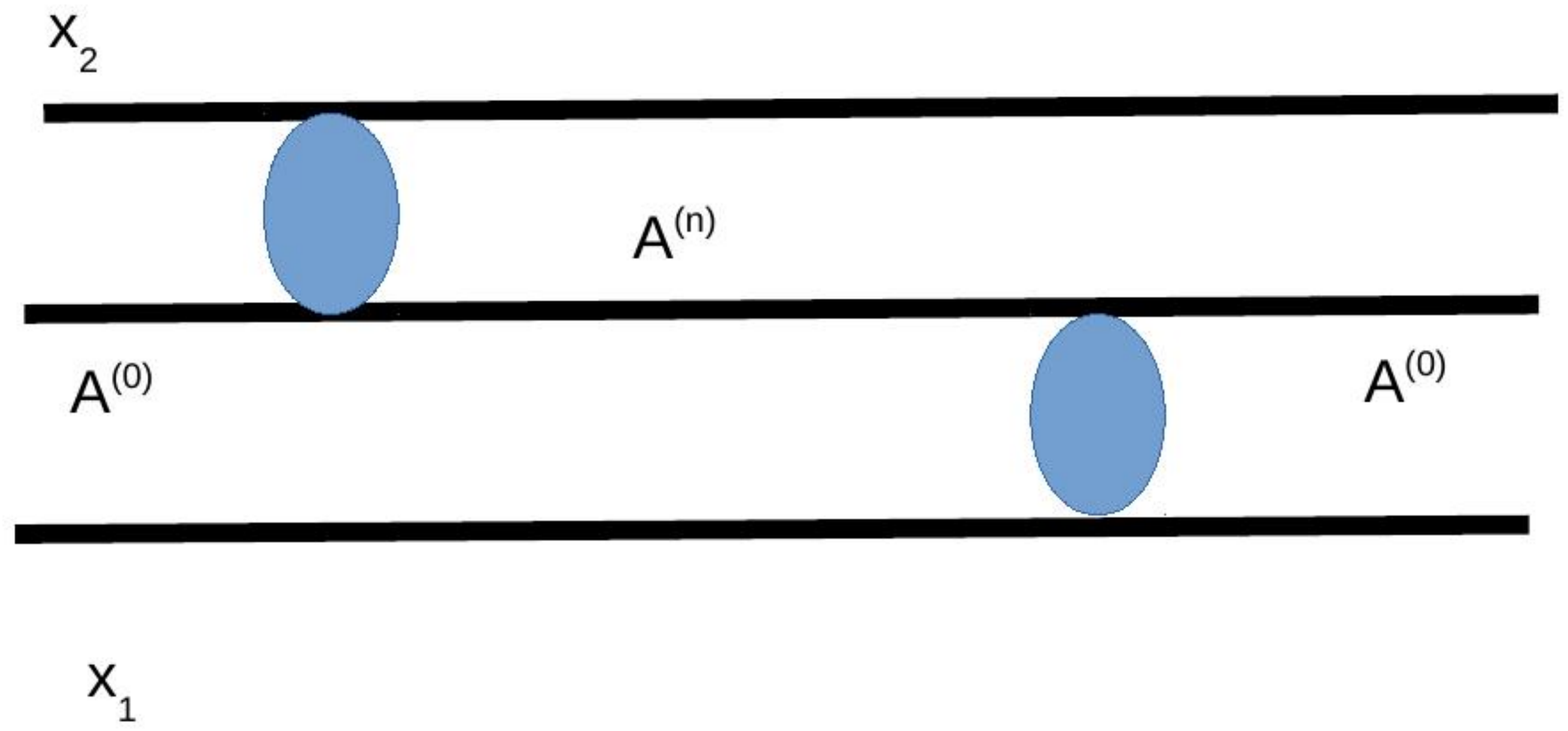,scale=0.4}} 
\vspace{-2cm}
\caption{Three-body optical potential $U_{3B}$. Excitation of the target by particle $x_1$ and 
de-excitation by $x_2$ (see Eq. (\ref{U3B})).}
\label{fig1}
\end{figure}

We remind the reader that $W_{3B}$ results from the average of processes involving the excitation of the target by one of the fragment and its de-excitation by the other fragment as illustrated in Fig. \ref{fig1}.

In the language of projection operators, the 3B optical potential, $U_{3B}$, whose imaginary part is -$W_{3B}$, is given by

\begin{equation}
U_{3B} = PV_{{x_1}A}Q (QG_{XA}(E_{x})Q) QV_{{x_2}A}P + PV_{{x_2}A}Q (QG_{XA}(E_x)Q) QV_{{x_1}A}P \label{U3B}
\end{equation}

where,  the Q-projected 3B Green's function, $QG_{XA}(E_x)Q\equiv QG_{XA}Q$, is given by
\begin{equation}
QG_{XA}Q = \frac{1}{E_x - QH_{0}Q + Q[V_{{x_1}A} + V_{{x_2}A}]PG_{0}P[V_{{x_1}A} + V_{{x_2}A}]Q + i\varepsilon}
\end{equation}
The imaginary part of $U_{3B}$ is now easily calculated. Since $[V_{{x_1}A}+V_{{x_2}A}]\,PG_{0}P\,[V_{{x_1}A}+V_{{x_2}A}]$ is non-Hermitian operator, the calculation proceeds as in Eq. (\ref{ImaG}),
\begin{multline}
Im[QG_{XA}Q] = -\pi \Omega^{(-)}_{Q}\delta(E_x - QH_{0}Q)(\Omega^{(-)}_{Q})^{\dagger} + \\  -(QG_{XA}Q)^{\dagger} Q[V_{{x_1}A} + V_{{x_2}A}]\,\delta(E_x- PH_{0}P)\,[V_{{x_2}A} + V_{{x_2A}}]QGQ   
\end{multline}

Thus,
\begin{multline}
W_{3B} = \pi[PV_{{x_1A}}Q \Omega^{(-)}_{Q}\delta(E_x - QH_{0}Q) (\Omega^{(-)}_{Q})^{\dagger}QV_{{x_2A}}P  +(x_{1}\leftrightarrow x_{2})+
\\
+ PV_{{x_1A}}Q (QGQ)^{\dagger} Q[V_{{x_1A}} + V_{{x_2A}}]P\delta(E_x - PH_{0}P)P[V_{{x_1A}}+V_{{x_2A}}]Q  (QGQ) QV_{{x_2A}}P+
\\
 + PV_{{x_2A}}Q (QGQ)^{\dagger} Q[V_{{x_1A}} + V_{{x_2A}}]P\delta(E_x - PH_{0})P)P[V_{{x_1A}} + V_{{x_2A}}]Q  (QGQ)  QV_{{x_1A}}P] 
\end{multline}

Therefore the reactive content of $W_{3B}$ is simple to discern. The first term corresponds to the already announced excitation of the target by one fragment followed by a de-excitation through the action of the second fragment. The last two terms corresponds to absorption of the two fragments by the target. There are eight  terms which describe the different ways this absorption is manifested. It is evident that a detailed evaluation of $U_{3B}$ is a formidable task. We venture to hypothesize the following. The first term above is replaced by a non-local separable term with an energy-dependent "strength'" such that the exitation-de-excitation term is represented by $F_{1}(r) f(E) F_{2}(r^{\prime})$. If the two fragments are identical, such as the case of two neutrons, then $F_{1} = F_{2}$. The second and third term, the "capture" part, is replaced by a single term which involves the reaction/fusion of two correlated fragments. The correlation is induced by the interaction $V_{{x_1}{x_2}}$, which besides scattering the two fragments, could bind them in a resonance or quasi-bound state. Accordingly we replace the very complicated structure above by a simple effective two-body fusion. This applies to the calculation of the reaction cross section, $\sigma^{3B}_{R}$.  Further study of this matter is in progress. \\

At this point it is important to mention that the calculation of \cite{Potel2015} is based on the prior form of the DWBA version of the, three-body, Austern cross section, while those of \cite{Moro2-2015}, and \cite{Carlson2015} are based on the post form of the DWBA version of that formula. The relation between these two versions were discussed in \cite{Ichimura1990, HFM1990}, and a brief account of derivation of this relation is presented in the appendix.

\section{Distorted wave  expansion of the four-body breakup cross section}\label{DWBA-4B}
The derivation of a DWBA three-body inclusive breakup cross section which results in the IAV, UT, and HM expressions, and the subsequent derivation of the relation between these cross sections, relied on the use of the dominant Faddeev component of the three-body wave function \cite{HFM1990}. To obtain a similar formulas in the four-body inclusive breakup cross section, we have to rely on the Faddeev-Yakubovsky decomposition \cite{YakSJP67}.

\subsection{The Faddeev-Yakubovsky equations}

The full scattering wave function
is reconstructed from the Faddeev components as:
\begin{equation}
|\Psi^{4B(+)}_0\rangle=\sum_{i>j} |F^{(+)}_{ij}\rangle
\end{equation}
where 
\begin{equation}
|F^{(+)}_{ij}\rangle=G^{(+)}_0v_{ij}|\Psi^{4B(+)}_0\rangle
\end{equation}
with  the two-body potential $v_{ij}$
 between the particles i and j, the free four-body resolvent
$G^{(+)}_0=\left[E-H_0+\imath \epsilon\right]^{-1}$ and  kinetic energy $H_0$.

These components are further decomposed in terms
of the  Faddeev-Yakubovsky (FY) amplitudes \cite{YakSJP67}, 
\begin{equation}
|F^{(+)}_{ij}\rangle=|K_{ij,k}^{l(+)}\rangle+|K_{ij,l}^{k(+)}\rangle+|H^{(+)}_{ij,kl}\rangle,
\label{fad}
\end{equation}
which are defined as
\begin{eqnarray}
|K_{ij,k}^{l(+)}\rangle=G^{(+)}_{ij}v_{ij}G^{(+)}_0(v_{ik}+v_{jk})|\Psi^{4B(+)}_0\rangle\;(i<j),
\label{K}
\\ \nonumber \\
|H_{ij,kl}^{(+)}\rangle=G^{(+)}_{ij}v_{ij}G^{(+)}_0v_{kl}|\Psi^{4B(+)}_0\rangle\;(i<j,k<l),
\label{H}
\end{eqnarray}
where $G^{(+)}_{ij}$ is two-body resolvent within the four-body system and system given by $G^{(+)}_{ij}=[E-H_0-v_{ij}+\imath\epsilon]^{-1}$.

The amplitude $K$ and $H$ may also be written in terms of
the Faddeev components  as:
\begin{eqnarray}
&&|K_{ij,k}^{l(+)}\rangle=G^{(+)}_{ij}v_{ij}(|F^{(+)}_{ik}\rangle+|F^{(+)}_{jk})\rangle)\;(i<j),
\label{k1}
\\ \nonumber \\
&&|H_{ij,kl}^{(+)}\rangle=G^{(+)}_{ij}v_{ij}|F^{(+)}_{kl}\rangle\;(i<j,k<l).
\label{h1}
\end{eqnarray}
Substituting Eq.  (\ref{fad})  in Eqs. (\ref{k1}) and  (\ref{h1})
and using that $G_{ij}^{(+)}v_{ij}=G_0^{(+)}t_{ij}$, and introducing the boundary condition we have that:
\begin{eqnarray}
\nonumber
|K_{ij,k}^{l(+)}\rangle=&&|\Psi^{3B}_{ij,k}\,;\,k_l\rangle+G_0^{(+)}\,t_{ij}(E-E_{ij,k}-E_l)[|K_{ik,j}^{l(+)}\rangle+
|K_{jk,i}^{l(+)}\rangle \\
&& +|K_{ik,l}^{j(+)}\rangle
+|K_{jk,l}^{i(+)}\rangle  +|H_{ik,jl}^{(+)}\rangle+|H_{jk,il}^{(+)}\rangle], \label{kf} \\
|H_{ij,kl}^{(+)}\rangle=&&G_0^{(+)}\,t_{ij}(E-E_{ij,kl}-E_{kl})[|K_{kl,i}^{j(+)}\rangle
+ |K_{kl,j}^{i(+)}\rangle +|H_{kl,ij}^{(+)}\rangle],\label{hf}
\end{eqnarray}
expressing the collision of an incoming bound three-body system $\{ijk\}$ with the target particle $l$. 
The  amplitude $|\Psi^{3B}_{ij,k}\rangle=G_0(E_B)\,v_{ij}|\Psi^{3B}_{ijk}\rangle$ is the Faddeev component of the three-body
bound state wave function of the projectile $|\Psi^{3B}_{ijk}\rangle$ composed by particles $\{ijk\}$. The relative momentum state of the projectile
to the target $(l)$ is $|\,k_l\rangle$.

\newpage

\subsection{Derivation of the CFH formula within the Faddeev-Yakubovsky Formalism}

The full scattering wave function
is a solution of the Lippman-Schwinger equation:
\begin{equation}
|\Psi^{4B(+)}_0\rangle=G^{(+)}_{x_1\,x_2\,A\,,b}\left(V_{x_1b}+V_{x_2b}\right)|\Psi^{4B(+)}_0\rangle
\end{equation}
where the Green's function,
\begin{equation}
G^{(+)}_{x_1\,x_2\,A\,,b}=\left[E-K_{x_1}-K_{x_2}-K_{b}-V_{x_1x_2}-U_{x_1A}-U_{x_2A}-U_{bA}+\imath\epsilon\right]^{-1}
\end{equation}
 also contains the optical potential $U_{bA}$ and can be expressed as:
\begin{equation}
G^{(+)}_{x_1\,x_2\,A\,,b}=G^{(+)}_{x_1\,x_2}+G^{(+)}_{x_1\,x_2\,A\,,b}\left(U_{x_1A}+U_{x_2A}+U_{bA}\right)G^{(+)}_{x_1\,x_2}
\end{equation}
and
\begin{equation}
|\Psi^{4B(+)}_0\rangle=G^{(+)}_{x_1\,x_2}\left(V_{x_1b}+V_{x_2b}\right)
|\Psi^{4B(+)}_0\rangle+G^{(+)}_{x_1\,x_2\,A\,,b}\left(U_{x_1A}+U_{x_2A}+U_{bA}\right)G^{(+)}_{x_1\,x_2}\left(V_{x_1b}+V_{x_2b}\right)|\Psi^{4B(+)}_0\rangle
\end{equation}
and using that:
\begin{equation}
G^{(+)}_{x_1\,x_2}=G^{(+)}_{0}+G^{(+)}_{x_1\,x_2}\,V_{x_1x_2}\,G^{(+)}_{0}
\end{equation}
we obtain:
\begin{multline}
|\Psi^{4B(+)}_0\rangle=
\left(1+G^{(+)}_{x_1\,x_2}\,V_{x_1x_2}\right)\,G^{(+)}_{0}\left(V_{x_1b}+V_{x_2b}+U_{bA}\right)|\Psi^{4B(+)}_0\rangle+\\
+G^{(+)}_{x_1\,x_2\,A\,,b}\left(U_{x_1A}+U_{x_2A}+U_{bA}\right)\left(1+G^{(+)}_{x_1\,x_2}\,V_{x_1x_2}\right)\,G^{(+)}_{0}\left(V_{x_1b}+V_{x_2b}\right)|
\Psi^{4B(+)}_0\rangle
\end{multline}
Introducing the definitions of the F-Y components, namely:
\begin{eqnarray}
&&|F^{(+)}_{x_1b}\rangle=\,G^{(+)}_{0}\,V_{x_1b}|\Psi^{4B(+)}_0\rangle\\
&&|F^{(+)}_{x_2b}\rangle=\,G^{(+)}_{0}\,V_{x_2b}|\Psi^{4B(+)}_0\rangle\\
&&|K^{A(+)}_{x_1x_2\,,b}\rangle=
G^{(+)}_{x_1\,x_2}\,V_{x_1x_2}\left(|F^{(+)}_{x_1b}\rangle+|F^{(+)}_{x_2b}\rangle\right)
\end{eqnarray}
we get that:
\begin{equation}
|\Psi^{4B(+)}_0\rangle=
\left[1+G^{(+)}_{x_1\,x_2\,A\,,b}\left(U_{x_1A}+U_{x_2A}+U_{bA}\right)\right]\left(|F^{(+)}_{x_1b}\rangle+|F^{(+)}_{x_2b}\rangle+|K^{A(+)}_{x_1x_2\,,b}\rangle\right)
\end{equation}
Using  the decomposition of definition of the Faddeev components in terms of the F-Y components:
\begin{eqnarray}
&&|F^{(+)}_{x_1b}\rangle=|K^{A(+)}_{x_1b\,,x_2}\rangle+ |K^{x_2(+)}_{x_1b\,,A}\rangle+ |H^{(+)}_{x_1b\,,x_2A}\rangle\\
&&|F^{(+)}_{x_2b}\rangle=|K^{A(+)}_{x_2b\,,x_1}\rangle+ |K^{x_1(+)}_{x_2b\,,A}\rangle+ |H^{(+)}_{x_2b\,,x_1A}\rangle\\
\end{eqnarray}
we find the form of the four-body scattering state
\begin{multline}
|\Psi^{4B(+)}_0\rangle=
\left[1+G^{(+)}_{x_1\,x_2\,A\,,b}\left(U_{x_1A}+U_{x_2A}+U_{bA}\right)\right]
\left(|K^{A(+)}_{x_1b\,,x_2}\rangle+ |K^{A(+)}_{x_2b\,,x_1}\rangle+
|K^{A(+)}_{x_1x_2\,,b}\rangle+\right.\\ \left.
+|K^{x_2(+)}_{x_1b\,,A}\rangle+
 |H^{(+)}_{x_1b\,,x_2A}\rangle
 + 
 |K^{x_1(+)}_{x_2b\,,A}\rangle+ |H^{(+)}_{x_2b\,,x_1A}\rangle
 \right)
\end{multline}
Isolating the first three F-Y components of the $K-$type, with $A$ as spectator, which seems to be
 dominant contribution to the four-body scattering, we make the following approximation:
\begin{equation}
|\Psi^{4B(+)}_0\rangle\approx\left[1+G^{(+)}_{x_1\,x_2\,A\,,b}\left(U_{x_1A}+U_{x_2A}+U_{bA}\right)\right]
\left(|K^{A(+)}_{x_1b\,,x_2}\rangle+ |K^{A(+)}_{x_2b\,,x_1}\rangle+
|K^{A(+)}_{x_1x_2\,,b}\rangle\right)\approx |\chi^{(+)}_a\,,\,\Phi_{x_1x_2b}\rangle
\end{equation}
where $\chi^{(+)}_A$ is the scattering wave of the C.M. of the three-body projectile and
$\Phi_{x_1x_2b}$ is the projectile internal wave function. Therefore, using this dominant F-Y components
of the four-body wave function, we can make the following approximation for the continuum four-body wave function as:
\begin{equation}
|\Psi^{4B(+)}_0\rangle=G^{(+)}_{x_1\,x_2\,A\,,b}\left(V_{x_1b}+V_{x_2b}\right)|\Psi^{4B(+)}_0\rangle
\approx\, G^{(+)}_{x_1\,x_2\,A\,,b}\left(V_{x_1b}+V_{x_2b}\right) |\chi^{(+)}_a\,,\,\Phi_{x_1x_2b}\rangle \label{psi4bapp}
\end{equation}

The relevant term in the inelastic breakup formula (\ref{CFH}) is
\begin{equation}
 \tau_X=\langle\Psi_{0}^{4B(+)}|\chi_{b}^{(-)}\rangle(W_{x_1} + W_{x_2} + W_{3B})\langle\chi_{b}^{(-)}|\Psi_{0}^{4B(+)}\rangle \label{CFH1}
\end{equation}
which will be manipulated with the help of our approximation of the four-body scattering wave function (\ref{psi4bapp}) as follows:
\begin{equation}
 \tau_X\approx\, \langle\chi^{(+)}_a\,,\,\Phi_{x_1x_2b}|\left(V_{x_1b}+V_{x_2b}\right)G^{(+)\,\dagger}_{x_1\,x_2\,A\,,b}|\chi_{b}^{(-)}\rangle(W_{x_1} + W_{x_2} + W_{3B})\langle\chi_{b}^{(-)}|G^{(+)}_{x_1\,x_2\,A\,,b}\left(V_{x_1b}+V_{x_2b}\right) |\chi^{(+)}_a\,,\,\Phi_{x_1x_2b}\rangle \label{psi4bapp-1}
\end{equation}
where we introduce the spectator distorted wave function as:
\begin{equation}
|\hat{\rho}^{CFH}_{b}\rangle = \langle\chi_{b}^{(-)}|G^{(+)}_{x_1\,x_2\,A\,,b}\left(V_{x_1b}+V_{x_2b}\right) |\chi^{(+)}_a\,,\,\Phi_{x_1x_2b}\rangle \label{psi4bapp-2}
\end{equation}
We can rewrite the above formula by using the eigenvalue equations:
\begin{eqnarray}
&&\left(E-K_b-K_{x_1}-K_{x_2}-V_{x_1b}-V_{x_2b}-V_{x_1x_2}-U_{aA}\right)|\chi^{(+)}_a\,,\,\Phi_{x_1x_2b}\rangle =0\label{psi4bapp-3}\\
&&\left(E_b-K_b-U^{\dagger}_{b\,A}\right)|\chi^{(-)}_b\rangle =0 \label{psi4bapp-4}
\end{eqnarray}
and manipulating this equation we identify that:
\begin{eqnarray}
\left(V_{x_1b}+V_{x_2b}\right)|\chi^{(+)}_a\,,\,\Phi_{x_1x_2b}\rangle =
\left(E-K_b-K_{x_1}-K_{x_2}-V_{x_1x_2}-U_{aA}\right)|\chi^{(+)}_a\,,\,\Phi_{x_1x_2b}\rangle \label{psi4bapp-5}
\end{eqnarray}
Substituting Eq. (\ref{psi4bapp-5}) in (\ref{psi4bapp-2}) we get
\begin{equation}
|\hat{\rho}^{CFH}_{b}\rangle = \langle\chi_{b}^{(-)}|G^{(+)}_{x_1\,x_2\,A\,,b}\left(E-K_b-K_{x_1}-K_{x_2}-V_{x_1x_2}
-U_{aA}\right)|\chi^{(+)}_a\,,\,\Phi_{x_1x_2b}\rangle  \label{psi4bapp-6}
\end{equation}
and reminding that $G^{(+)}_{x_1\,x_2\,A\,,b}$ contains the optical potential of $b$ with respect to $A$
\begin{eqnarray}\label{rhocfh1}
|\hat{\rho}^{CFH}_{b}\rangle  &=&\langle\chi_{b}^{(-)}|G^{(+)}_{x_1\,x_2\,A\,,b}\left(E-K_b-K_{x_1}-K_{x_2}-V_{x_1x_2}
-U_{aA}\right)|\chi^{(+)}_a\,,\,\Phi_{x_1x_2b}\rangle \nonumber\\
&=&G^{(+)}_{x_1\,x_2\,A}(E-E_b)\langle\chi_{b}^{(-)}|\left(E-K_b-K_{x_1}-K_{x_2}-V_{x_1x_2}-U_{aA}\right)
|\chi^{(+)}_a\,,\,\Phi_{x_1x_2b}\rangle
\end{eqnarray}
where
\begin{equation}
G^{(+)}_{x_1\,x_2\,A}(E-E_b)=\left[E-E_b-K_{x_1}-K_{x_2}-V_{x_1x_2}-U_{x_1A}-U_{x_2A}+\imath\epsilon\right]^{-1}
\end{equation}
and using once more Eq.(\ref{psi4bapp-4}), one gets from (\ref{rhocfh1}) that:
\begin{eqnarray}\label{rhocfh2}
|\hat{\rho}^{CFH}_{b}\rangle &=&G^{(+)}_{x_1\,x_2\,A}(E-E_b)\langle\chi_{b}^{(-)}|\left(E-E_b-K_{x_1}-K_{x_2}-V_{x_1x_2}+U_{bA}-U_{aA}\right)
|\chi^{(+)}_a\,,\,\Phi_{x_1x_2b}\rangle \nonumber\\
&=&G^{(+)}_{x_1\,x_2\,A}(E-E_b)\langle\chi_{b}^{(-)}|\left(E-E_b-K_{x_1}-K_{x_2}-V_{x_1x_2}-U_{x_1A}-U_{x_2A}\right)
|\chi^{(+)}_a\,,\,\Phi_{x_1x_2b}\rangle \nonumber\\
&&+G^{(+)}_{x_1\,x_2\,A}(E-E_b)\langle\chi_{b}^{(-)}|\left(U_{x_1A}+U_{x_2A}+U_{bA}-U_{aA}\right)
|\chi^{(+)}_a\,,\,\Phi_{x_1x_2b}\rangle
 \label{psi4bapp-7} 
\end{eqnarray}
and finally:
\begin{eqnarray}
|\hat{\rho}^{CFH}_{b}\rangle &=&\langle\chi_{b}^{(-)}|\chi^{(+)}_a\,,\,\Phi_{x_1x_2b}\rangle 
+G^{(+)}_{x_1\,x_2\,A}(E-E_b)\langle\chi_{b}^{(-)}|\left(U_{x_1A}+U_{x_2A}+U_{bA}-U_{aA}\right)
|\chi^{(+)}_a\,,\,\Phi_{x_1x_2b}\rangle
 \label{psi4bapp-8} 
\end{eqnarray}
which generalizes our previous formula for a two-body composite projectile \cite{HFM1990} for a three-body projectile. The
Ichimura, Austern and Vincent formula extended to the four-body situation, with a three-body projectile plus the target: 
\begin{eqnarray}
|\hat{\rho}^{CFH}_{b}\rangle &=&|\hat{\rho}^{4B}_{HM}\rangle + |\hat{\rho}^{4B}_{UT}\rangle
 \label{psi4bapp-9} 
\end{eqnarray}
and introducing HM and UT at the four-body level we write that:
\begin{equation}
|\hat{\rho}^{4B}_{HM}\rangle =\langle\chi_{b}^{(-)}|\chi^{(+)}_a\,,\,\Phi_{x_1x_2b}\rangle 
\label{psi4bapp-10} 
\end{equation}
and
\begin{equation}
|\hat{\rho}^{4B}_{UT}\rangle=G^{(+)}_{x_1\,x_2\,A}(E-E_b)\langle\chi_{b}^{(-)}|\left(U_{x_1A}+U_{x_2A}+U_{bA}-U_{aA}\right)
|\chi^{(+)}_a\,,\,\Phi_{x_1x_2b}\rangle
 \label{psi4bapp-11} 
\end{equation}
these source functions are used in the cross-section formula (\ref{CFH-R}) to separate out the HM and UT contributions in addition to
a interference term.\\

The cross sections, $\sigma_{R}^{x_1}$, $\sigma_{R}^{x_2}$, of Eq. (\ref{CFH-R}) are related to the IAV cross section of the three-body theory through a convolution of the latter with the distorted wave densities, namely, $|\chi^{(+)}_{{x}_2}(\textbf{r}_{x_2})|^2$, 
and $|\chi^{(+)}_{{x}_1}(\textbf{r}_{x_1})|^2$, respectively. This can be easily seen if an eikonal-type approximation of the projectile distorted wave, $\chi^{(+)}_a = \chi^{(+)}_{b} \chi^{(+)}_{{x}_1}\chi^{(+)}_{{x}_2}$ is used. These distorted wave densities come from the solutions of a non-Hermitian Schr\"{o}dinger equation with the respective optical potentials. In Ref. \cite{Hussein1987}, these distorted wave densities were calculated and found to be related to the reaction cross section of $x_1$ and of $x_2$. To be more specific, we first write the source function $\hat{\rho}^{4B}_{HM}$, Eq. (\ref{psi4bapp-10})

\begin{equation}
\langle {\bf r}_{{x}_1}, {\bf r}_{{x}_2}|\hat{\rho}^{4B}_{HM}\rangle = \int d{\bf r}_{b} \Phi_{x_1x_2b}(\bf{r}_{{x}_1}, \bf{r}_{{x}_2}, \bf{r}_b) \langle\chi^{(-)}_{b}|\chi^{(+)}_{b}\rangle(\bf{r}_b) \chi^{(+)}_{{x}_1}(\bf{r}_{{x}_1})\chi^{(+)}_{{x}_2}(\bf{r}_{{x}_2})
\end{equation}
The overlap function $\langle\chi^{(-)}_{b}|\chi^{(+)}_{b}\rangle(\bf{r}_b) $ is the integrand of the elastic S-matrix element of the spectator fragment, $b$, $S_{\textbf{k}_{b}^{\prime}, \textbf{k}_{b}}=\int d\textbf{r}_{b} \langle\chi^{(-)}_{b}|\chi^{(+)}_{b}\rangle(\bf{r}_b)$.  Thus, we introduce the internal motion modified S-matrix of the b fragment, $\hat{S}_{b}(\bf{r}_{{x}_1}, \textbf{r}_{{x}_2}) \equiv \int d\bf{r}_{b} \Phi_{x_1x_2b}(\bf{r}_{{x}_1}, \bf{r}_{{x}_2}, \bf{r}_b)\langle\chi^{(-)}_{b}|\chi^{(+)}_{b}\rangle(\bf{r}_b) $. This then allows writing the source function as,

\begin{equation}
\langle \bf{r}_{{x}_1}, \bf{r}_{{x}_2}|\hat{\rho}^{4B}_{HM}\rangle = \hat{S}_{b}(\bf{r}_{{x}_1}, \textbf{r}_{{x}_2}) \chi^{(+)}_{{x}_1}(\bf{r}_{{x}_1})\chi^{(+)}_{{x}_2}(\bf{r}_{{x}_2})
\end{equation}

The cross section, $\sigma_{R}^{x_1}$, Eq. (\ref{sigma_1}), becomes,

\begin{equation}
\frac{E_{{x}_1}}{k_{{x}_1}}\sigma_{R}^{x_1} = \int d\bf{r}_{{x}_2}|\chi^{(+)}_{{x}_2}(\textbf{r}_{x_2})|^{2} \int d\bf{r}_{{x}_1}|\hat{S}_{b}(\bf{r}_{{x}_1}, \textbf{r}_{{x}_2})|^{2}\left[W(\bf{r}_{{x}_1})|\chi^{(+)}_{{x}_1}(\textbf{r}_{x_1})|^{2}\right]\label{psi4bapp-13}
\end{equation}

to be compared to the two-body reaction cross section of the fragment x, in the breakup of a two-cluster projectile, a = x + b,
\begin{equation}
\frac{E_{x}}{k_x}\sigma_{R}^{x} = \int d\bf{r}_{x} |\hat{S}_{b}(\bf{r}_{x})|^{2}
\left[W(\bf{r}_x)|\chi^{(+)}_{x}(\textbf{r}_{x})|^2\right] \label{psi3app-13}
\end{equation}
where,
\begin{equation}
\hat{S}_{b}(\bf{r}_{x}) = \int d\bf{r}_{b} \Phi_{xb}(\bf{r}_x, \bf{r}_b)\langle \chi^{(-)}_{b}|\chi^{(+)}_{b}\rangle (\bf{r}_{b})
\end{equation}
One sees clearly that the 4B cross section is damped compared to the 3B one owing, among other factors, to the presence of the distorted 
wave  density $|\chi^{(+)}_{{x}_2}(\textbf{r}_{{x}_2})|^2$ in the former.  This density is a decaying function of $\bf{r}_{{x}_2}$ owing to absorption described by the imaginary part of the optical potential $U_{{{x}_2}A}$. In fact, the inner product $\langle\chi_{{{x}_2}, \bf{k}^{\prime}}^{(+)}|\chi_{{{x}_2}, \bf{k}}^{(+)}\rangle \neq (2\pi)^{3}\delta (\bf{k}^{\prime} - \bf{k})$, exactly because of absorption. 
The orthonormality is recovered if the dual state of $|\chi^{(+)}\rangle$, $|\hat{\chi}^{(+)}\rangle$ is used, $\langle\hat{\chi}_{{{x}_2}, \bf{k}^{\prime}}^{(+)}|
\chi_{{{x}_2}, \bf{k}}^{(+)}\rangle = (2\pi)^{3}\delta (\bf{k}^{\prime} - \bf{k})$. The dual state is the solution of scattering equation with the optical potential replaced by its complex conjugate \cite{Hussein1987}.

In addition, the cross-section formula (\ref{psi4bapp-13}) applied to a three-body halo nuclei, e.g., an weakly bound 
neutron-neutron-core nuclei close to the drip-line, has built in a long-range correlation between the two neutrons and the
core $b$, through the extended wave function of the halo and $|\chi^{(+)}_{{x}_2}(\textbf{r}_{{x}_2})|^2$.

The same considerations can be made with respect to the cross section $\sigma_{R}^{x_2}$. Similar observations can be made if the IAV or UT expressions were used. Finally, it is instructive to compare the above cross sections to the "free" one, where x is the primary projectile,
\begin{equation}
\frac{E}{k}\sigma_{R} = \int {\bf dr} \left[W({\bf r}) |\chi^{(+)}(\bf{r})|^2\right] \, .
\end{equation}

An interesting limit to consider is the Serber model \cite{Serber1950}. If one sets $\langle\chi^{(-)}_{b}|\chi^{(+)}_{b}\rangle(\bf{r}_b)  = 1$, and replace the projectile intrinsic wave function by its Fourier transform and take it outside the integral, then, Eq.(\ref{psi3app-13}), becomes, $\sigma_{R}^{x} \varpropto |\Phi_{xb}(q_{xb})|^2\sigma_{R}(E_x)$. The inclusive non-elastic breakup cross section, is then reduced to the Serber cross section \cite{Serber1950},

\begin{equation}
\frac{d^{2}\sigma_{b}}{d\Omega_{b}dE_{b}} = \rho_{b}(E_{b}) |\Phi_{xb}(\textbf{q}_{xb})|^2 \sigma_{R}(E_x)\, .
\end{equation}

\section{Discussion and Conclusions}
In this paper we have derived the 3-fragment projectile inclusive breakup cross section and pointed out the major differences from the corresponding cross section in the case of two-fragment projectile currently used in the calculations. Our theory permits the study of fragment-fragment correlation through a judicial coincidence measurement of the elastic breakup part of the cross section. The inclusive non-elastic breakup, or incomplete fusion, part of the cross section we have derived formulae reminiscent of the so-called Austern formula, with two major differences. Our 4-body formula contains reference to the three-body nature of the fusing two fragments and to the intrinsically three -body "direct" process, which permits the excitation of the target by one of the fragments followed by the target de-excitation by the other fragment. We called the 4-body formula the CFH expression. The imaginary part of the optical potential is found to be composed by the sum of two one-fragment potentials, plus a new, 3-body part, which contains the fusion of the two fragments. We proposed a simplified model to deal with this three-body absorption term in the imaginary part of this latter potential. Our results should be quite useful in the study of inclusive breakup of Borromean nuclei, where two neutrons are involved in the reaction mechanism. Hybrid theories, such as the Surrogate Method, can now be extended to the case of, say, tritium breakup. The DWBA version of the theory is also developed. Such a distorted wave approximation required the employment of the four-body Faddeev-Yakubovsky equations \cite{Faddeev1961, YakSJP67}, just as the three-body theory requiring the three-body Faddeev equations \cite{HFM1990, Faddeev1961}. This has been accomplished in the previous section and the results confirms that the general structure of the cross section is similar to the three-body case, with the full post form or the four-body IAV one can be written as the sum of the prior UT cross section plus the four-body HM one plus the interference term. The major difference between the 4B and 3B cases resides in the structure of the reaction cross sections for the absorption of one of the interacting fragments, which we found to be damped by the absorption effect of the other fragment. In general we expect that in cases such as (t,p), the one neutron absorption cross section is smaller than the  corresponding one in the (d,p) reaction.\\

{\it Acknowledgements.} 
This work was partly supported by the Brazilian agencies, Funda\c c\~ao de Amparo \`a Pesquisa do Estado de
 S\~ao Paulo(FAPESP), the 
Conselho Nacional de Desenvolvimento Cient\'ifico e Tecnol\'ogico  (CNPq). MSH also acknowledges a Senior Visiting Professorship granted by the Coordena\c c\~ao de Aperfei\c coamento de Pessoal de N\'ivel Superior (CAPES), through the CAPES/ITA-PVS program.

\newpage
{\Large Appendix}\\

\begin{enumerate}

\item \textbf{The derivation of Eq.(\ref{Goptical}) using the projection operator techniques};\\

{If $P = |\Phi^{0}_{A}\rangle \langle \Phi^{0}_{A}|$, and $Q = 1 - P$, then the average three-body Green's function is $PG^{(+)}_{{x_1}{x_2}A}P$. Using the Lippmann-Schwinger equation for $G^{(+)}_{{x_1}{x_2}A} = G^{(+)}_{{x_1}{x_1},0} + G^{(+)}_{{x_1}{x_1}, 0}[V_{{x_1}A} + V_{{x_2}A}] G^{(+)}_{{x_1}{x_2}A}$, with $G^{(+)}_{{x_1}{x_1}, 0} = [E_x - K_{x_1} - K_{x_2} - V_{{x_2}{x_2}} + i\varepsilon]^{-1}$ which does not couple $P$ to $Q$, namely, $PG^{(+)}_{{{x_1}{x_1}}, 0}Q = 0$. Then, simplifying the notation, $G = G_{0} + G_{0}VG$, and $G = [E_x - H +i\varepsilon]^{-1}$ we have after P- and Q- projecting
\begin{equation}
PGP = PG_{0}P + PG_{0}P + PG_{0}P (PVP) PGP  + PG_{0}P (PVQ) QGP 
\end{equation}
and
\begin{equation}
QGP = QG_{0}Q (QVQ) QGP + QG_{0}Q (QVP) PGP 
\end{equation}
The above two equations can be solved for $PGP$ using well known manipulations and the final solution is
\begin{equation}
PGP = \frac{1}{E_x - PHP - PVQ G_{Q} QVP + i\varepsilon} \equiv \frac{1}{E_x - H_0 - PVP - PVQ G_{Q}QVP + i\varepsilon}
\end{equation}
where $G_{Q} \equiv [E_x - QHQ + i\varepsilon]^{-1}$}.\\
The structure of $PGP$ above is well known and we write it here to make our paper as self contained as possible. It

\item\textbf{The imaginary part of a resolvent with non-hermitian terms is calculated as follows}\\

{Write $G_{x} = \frac{1}{A + B}$ where $A$ is Hermitian and $B$ is not. Then
$G^{-1} = A + B +i\varepsilon $ and $(G^{-1})^{\dagger} = A + B^{\dagger} - i\varepsilon $. The 
difference $G^{-1} - (G^{-1})^{\dagger} = 2i\varepsilon + B - B^{\dagger}$. The factor 
$2i\varepsilon$ can be replaced by $G_{A}^{-1} - (G_{A}^{-1})^{\dagger}$, where $G_{A} = [A+ 
i\varepsilon]^{-1}$. Thus $G^{-1} - (G^{-1})^{\dagger} = 
G_{A}^{-1} - (G_{A}^{-1})^{\dagger} + B - B^{\dagger}$.  Multiplying the last equation from the 
right by $G$ and from the left by $G^{\dagger}$ we get $G^{\dagger} - G = -2i\pi[1 + 
G^{\dagger}B^{\dagger}]\delta(A)[1 + BG] + G^{\dagger}[B - B^{\dagger}]G$, where we have 
used 
$G = 
G_{A} + G_{A} B G = G_{A} [1 + BG]$, and $G^{\dagger} = [1 + G^{\dagger} 
U^{\dagger}]G_{A}^{\dagger}$. 
Thus,
\begin{equation}
G - G^{\dagger} = -2i\pi \Omega^{(-)}\delta(A)(\Omega^{(-)})^{\dagger} + (G^{(+)})^{\dagger}[B - 
B^{\dagger}]G^{(+)} \label{ImG}
\end{equation}
The M\"oller operator is $\Omega^{(-)} = [1 + (G^{(+)})^{\dagger}U^{\dagger}]$. When operating on 
the eigenfunction of $A$ it generates a distorted wave}.\\

Thus Im$G_{x}^{(+)} = Im [E_x - K_x - U_x + i\varepsilon]^{-1}$ can be calculated easily. We use the 
identity derived above, Eq.(\ref{ImG}).,
\begin{equation}
Im G_{x}^{(+)} = -\pi \Omega^{(-)}_{x} \delta (E_x - K_x)(\Omega^{(-)}_{x})^{\dagger} - 
(G^{(+)}_{x})^{\dagger}W_{x}G^{(+)}_{x}
\end{equation}

where, $\Omega^{(-)}_x= (1 + G^{(-)}_{x}U_{x}^{\dagger})$ is the M\"oller operator, and $ -W_x$ is  
the imaginary part of the $x$ optical potential, $U_x = ReU_x - iW_x$.\\

\item \textbf{The structure of the cross section in the two-fragment projectile case $a = b + x$}\\

With the formula for Im $G_{x}^{(+)}$ derived above, the cross section then becomes,
\begin{equation}
\frac{d^{2}\sigma_b}{dE_{b}d\Omega_{b}} = \frac{d^{2}\sigma^{(EB)}_b}{dE_{b}d\Omega_{b}} + 
\frac{d^{2}\sigma^{(INEB)}_b}{dE_{b}d\Omega_{b}}
\end{equation}
where the elastic breakup cross section is given by,
\begin{equation}
\frac{d^{2}\sigma^{(EB)}_b}{dE_{b}d\Omega_{b}} = \frac{2\pi}{\hbar v_a}\rho_{b}
(E_b)\sum_{\textbf{k}_x} \left|\langle \chi^{(-)}_{x}\chi^{(-)}_{b}|V_{bx}|\Psi_{0}^{(+)}\rangle 
\right|^2 \delta(E - E_b - E_{{\textbf{k}_x}})
\end{equation}
and the inclusive non-elastic breakup (INEB)
\begin{equation}
\frac{d^{2}\sigma^{(INEB)}_b}{dE_{b}d\Omega_{b}} = -\frac{2}{\hbar v_a}\rho_{b}(E_b) 
\left\langle\Psi_{0}^{(+)}\left|V_{bx}\right|\chi^{(-)}_{b})(G^{(+)}_{x})^{\dagger}W_{x}G^{(+)}_{x}
(\chi^{(-)}_{b}\left|V_{bx}\right|\Psi_{0}^{(+)}\right\rangle =
 \end{equation}

Writing $V_{xb} = [V_{xb} + K_b + K_x + U_x +U_b -E] - [K_b +K_x + U_b + U_x -E]$ and using $[V_{xb} 
+ K_b + K_x + U_x +U_b -E]|\Psi_{0}^{(+)}\rangle = 0$, and further using $-G_{x}
(\chi^{(-)}_{b})^{\star}[K_b +K_x + U_b + U_x -E] = - G_{x}^{(+)}[K_x + U_x -E_x] 
(\chi^{(-)}_{b})^{\star} = - (\chi_{b}^{(-)})^{\star}$, we obtain the Austern equation, Eqs. 
(\ref{IC, exact}, \ref{NEIB}, \ref{source-exact}).

The double differential cross section in the post representation
has been derived by \cite{Austern1987}, and further analyzed in Refs.  \cite{Ichimura1990, HFM1990}. 
After a lengthy formal manipulations the resulting double differential cross section  of the 
spectator particle $b$, is
\begin{equation}
\frac{d^{2}\sigma}{d\Omega_{b}dE_{b}} = \frac{d^{2}\sigma_{EB}}{d\Omega_{b}dE_{b}}  - \frac{2}{\hbar 
v_a} \rho_{b}(E_b) \left<\hat{\rho}_{x, exact}\big|W_{x}\big|\hat{\rho}_{x, exact}\right> \label{IC, 
exact} ,
\end{equation}
Thus the inclusive non-elastic breakup cross section is
\begin{equation}
\frac{d^{2}\sigma_{NEIB}}{d\Omega_{b}dE_{b}} =  - \frac{2}{\hbar v_a} \rho_{b}(E_b) 
\left<\hat{\rho}_{x, exact}\big|W_{x}\big|\hat{\rho}_{x, exact}\right> \label{NEIB}
\end{equation}
where the x-fragment source function $\hat{\rho}_{x, exact}$ is given by
\begin{equation}
\hat{\rho}_{x. exact}(\bf{r}_x) = \left(\chi_{b}^{(-)}|\Psi^{(+)}\right> (\bf{r}_x) = \int 
d\bf{r}_{b} \left(\chi^{(-)}(\bf{r}_b)\right)^{\star}\Psi^{(+)}_{0}(\bf{r}_b, \bf{r}_x) 
\label{source-exact} 
\end{equation}

\item\textbf{The IAV, TU and HM cross sections in the three-body breakup case}\\

Remember that the wave function $|\Psi_{0}^{(+)}\rangle$ is the exact three-body wave function in 
the incident channel. To proceed further we use a DWBA approximation for this wave function, $|
\Psi_{0}^{(+)}\rangle$. This requires a careful analysis of the three-body scattering problem and 
one obtains,
\begin{equation}
|\Psi_{0}^{(+)}\rangle \approx G^{(+)}_{x,b}V_{xb}|\chi^{(+)}_{a}\phi_{a}\rangle
\end{equation}
where $G^{(+)}_{x,b}$ is the full Green's function $[E - K_b - K_x -U_b -U_x + i\varepsilon]^{-1}$. 

Using operator identities, one can derive
\begin{multline}
(\chi^{(-)}_b|\Psi_{0}^{(+)}\rangle  \approx (\chi^{(-)}_{b}|G^{(+)}_{x,b}V_{xb}|
\chi^{(+)}_{a}\phi_{a}\rangle\\
= G^{(+)}_{x}(\chi^{(-)}_{b}|(E - K_x -K_b - U_{a})|\chi^{(+)}_{a}\phi_{a}\rangle  \\
= G^{(+)}_{x}(\chi^{(+)}_{b}|(E - K_x + U_{b} - U_{a})|\chi^{(+)}_{a}\phi_{a}\rangle \\
= G^{(+)}_{x} (\chi^{(-)}_{b}|(U_{x}+ U_{b} - U_{a})|\chi^{(-)}\phi_{a}\rangle + G^{(+)}_{x}
(\chi^{(-)}_{b}|(E_x - K_x - U_{x})|\chi^{(+)}_{a}\phi_{a}\rangle \\
= G^{(+)}_{x} (\chi^{(-)}_{b}|(U_{x}+ U_{b} - U_{a})|\chi^{(+)}\phi_{a}\rangle + (\chi^{(+)}_{b}|
\chi^{(+)}\phi_{a}\rangle 
\end{multline}

Thus we obtain the desired relation,
\begin{equation}
G^{(+)}_{x}(\chi^{(-)}_b|V_{xb}|\chi^{(+)}_{a}\phi_a\rangle = G^{(+)}_{x} (\chi^{(-)}_{b}|(U_{x}+ U_{b} - U_{a})|\chi^{(+)}_{a}\phi_{a}\rangle + (\chi^{(-)}_{b}|\chi^{(+)}_{a}\phi_{a}\rangle 
\end{equation}
or,
\begin{equation}
\hat{\rho}_{x, IAV}(\bf {r}_x) = \hat{\rho}_{x, UT}(\bf {r}_x) + \hat{\rho}_{x, HM}(\bf {r}_x) 
\end{equation}
\end{enumerate}

The inclusive non-elastic breakup cross section then acquires the form,

\begin{equation}
\frac{d^{2}\sigma^{(IAV)}_b}{dE_{b}d\Omega_{b}} = \frac{d^{2}\sigma^{(UT)}_b}{dE_{b}d\Omega_{b}} + \frac{d^{2}\sigma^{(HM)}_b}{dE_{b}d\Omega_{b}} + \frac{d^{2}\sigma^{(INT)}_b}{dE_{b}d\Omega_{b}}
\end{equation}
where $\frac{d^{2}\sigma^{(INT)}_b}{dE_{b}d\Omega_{b}}$ is the interference contribution proportional to 2$Re\langle\hat{\rho}_{x, UT}|W_{x}|\hat{\rho}_{x, HM}\rangle$.

\end{document}